Estimation of Infection Rate and Prediction of Initial Infected Individuals of COVID-19

Seo Yoon Chae[1], Kyoung-Eun Lee[2,3], Hyun Min Lee[1], Nam Jung[1], Quang Ahn Le[1], Biseko Juma Mafwele[1], Tae Ho Lee[1], Doo Hwan Kim[1], and Jae Woo Lee[1,4,5]

[1]Department of Physics, Inha University, 100 Inharo Michuhol-gu, 22212 Republic of Korea

[2]Ecology and Future Research Institute, 21 Dusilo, Geumjeong-gu, Busan, Korea

[3]National Institute of Ecology, 1210 Geumgang-ro, Maseru-myeon, Seocheon-gun 33657, Korea

[4]Institue of Natural Basic Sciences, Inha University, 100 Inharo Michuhol-gu, 22212 Republic of Korea

[5]Institue of Advanced Computational Sciences, Inha University, 100 Inharo Michuhol-gu, 22212 Republic of Korea

Abstract

We consider the pandemic spreading of COVID-19 for some selected countries after the outbreak of the coronavirus in Wuhan City, China. We estimated the infection rate and the initial infecting individuals of COVID-19 by using the officially reported data at the early stage of the epidemic for the susceptible (S), infectable (I), quarantined (Q), and confirmed recovered ($R_k$) population model, so called SIQR$_k$ model. In the officially reported data we know the quarantined cases and the official reported recovered cases. We can't know the recovered cases from the asymptomatic cases. In the SIQR$_k$ model we can estimated the parameters and the initial infecting cases (confirmed cases + asymptomatic cases) from data fits. We obtained the infection rate in the range $\beta = 0.233 \sim 0.462$, the basic reproduction number $R_o = 1.8$~$3.5$, and the initial number of infected individuals $I(0) = 10$~$8409$ for some selected countries. By using fitting parameters we estimated that maximum time of the infection is around 50 days in Germany when the government are performing the quarantine policy. The disease is undergoing to a calm state about 6 months after the first patients were found.

I. Introduction

On December 31, 2019 Chinese authorities reported pneumonia cases from an unknown cause to World Health Organization (WHO) in Wuhan City, Hubei province, China. On January 7, 2020, the disease was identified as a new coronavirus first referred as 2019-nCov (SARS-CoV-2) and then named COVID-19. On January 11, 2020, China reported the first death by the coronavirus [1]. The victim was as 61-year old man in Wuhan. On January 20, 2020, a WHO reported the first confirmed cases outside China in Thailand, Japan, and South Korea [1]. The disease was spreading rapidly at Wuhan City and the cases were reported in outside Wuhan City. On 23 January 2020, China placed Wuhan, a city of 11 million people, under quarantine orders. All transportation departures were cancelled and suspended [1]. The president of WHO had declared the pandemic of COVID-19 in March 11, 2020. After the first report in December 31, 2019 at Wuhan City, Hubei province, China, the COVID-19 was spreading very quickly all over the world [2] and the COVID-19 is first pandemic disease in the 21 century.

Some countries such as Republic of Korea, Taiwan, Singapore, Chinese Hong Kong are controlling the disease successfully up to now. However, other countries like USA, Italy, Spain, France, UK etc. are suffering the outbreak and the shortages of the medical materials and hospitals. After the outbreak the scientists all over the world are struggling to find out the vaccine and the treating drug. In the highly connected societies the information and data for the diseases are shared through the internet, social media, and mass media. We can obtain information from the website such as worldometer [3] or livecornamap in South

Korea [4].

The flooding articles and preprints are appearing on the many journals and preprint websites. Recently the preprint websites like arXiv.org [5], bioRxiv [6], and medRxiv [7] are servicing the section for COVID-19 quick links. It is important to predict the spreading of the disease in the early stage of the outbreak. Many epidemic models are proposed based on the dynamic spreading models, agent-based models, Monte Carlo model, and data-based spreading models [9-14].

The evolution of the virus was described by the modified susceptible (S), infectious (I), recovered (R) population, so called SIR model. The prediction of COVID-19 evolution in Brazil has been suggested by the susceptible, infectious, quarantined, recovered (SIQR) model [9]. By the numerical analysis, he estimated the basis reproduction number $R_o = 5.25$ and the double time to be 2.72 days. The SIQR model includes a rate that quantifies the recovering of asymptomatic individuals for the evolution equation of the infection and recovering population. Peng *et al* introduced epidemic model of COVID-19 including the exposed population. Their model is called generalized SEIR model [10]. They introduced the time dependent parameters such as mortality rate and protection rate. Carcione *et al* reported the simulation results of the COVID-19 epidemic by the SEIR model [11]. They applied the model to the situation in the Italian Region of Lombardy. They estimated the basic reproduction number $R_o = 2.6$ in early stage of the outbreak. Fanelli and Piazza analyze and forecast the COVID-19 spreading by the SIRD model in China, Italy and France [8]. Padersen and Meneghini quantified the undetected COVID-19 cases and effects of containment measure in Italy. They introduced SIQR model [12] which include the rate for a patient to become non-infectious.

In this article we consider the $SIQR_K$ model based on only known data for the active cases and the recovered cases. We estimated the parameters of the model from the data of the reported cases for some selected countries. We obtained the infection rate and the initial number of the infected individuals. From the fitting parameters we estimated the basis reproduction number. We predict the maximum time of the infection and the annihilating period of the disease.

II. Epidemic Model

We consider an epidemic model of COVID-19 which is characterized by the variables $\{S(t), I(t), Q(t), R(t)\}$ denoting the susceptible population, infected population, quarantined population, and recovered population at time $t$. The total number of the population is satisfied a constraint $N = S(t) + I(t) + Q(t) + R(t)$ where $N$ is the total number of the population. Let us define the recovered population as $R(t) = R_k(t) + R_a(t)$ where $R_k(t)$ is known or confirmed recovered populations who are reported officially and $R_a(t)$ is the unknown or asymptomatic recovered populations who are infected, but do not show any symptoms. Under the homogenous mixing postulate, we consider a model, so called, SIQR$_K$ model as

$$\frac{dS(t)}{dt} = -\beta \frac{S(t)I(t)}{N}, \tag{1}$$

$$\frac{dI(t)}{dt} = \beta \frac{S(t)I(t)}{N} - (\alpha + \eta)I(t), \tag{2}$$

$$\frac{dQ(t)}{dt} = \eta I(t) - \gamma Q(t), \tag{3}$$

$$\frac{dR_k(t)}{dt} = \gamma Q(t). \tag{4}$$

$$\frac{dR_a(t)}{dt} = \alpha I(t). \tag{5}$$

In this model the parameter $\beta$ denotes the infection rate, the rate $\alpha$ with which patients become non-infectious by the recovering without any symptoms. The parameter $\eta$ is the rate of detection of new infecting people, and $\gamma$ is the recovering rate from the quarantined cases. In the SIQR$_k$ model, the infected populations are divided by the officially confirmed cases and the asymptomatic cases. We only know the official quarantined cases and the official recovered cases from the infection. We don't know the actual number of the infecting population due to the asymptomatic cases. The asymptomatic individuals are recovered without any severe suffering of the disease. The recovering population without symptoms are represented by $R_a(t)$. We propose the parameters included in the dynamic equations and the initial number of infecting cases which is sum of the officially known cases and the unknown population of the asymptomatic cases.

Table 1. Date of first case, state, and location of COVID-19 outbreaks for selected countries. The ranking is based on the total reported cases of the infection [3].

| Ranking | Country | First Report | State | Location |
|---|---|---|---|---|
| 1 | USA | Jan. 21 | Washington | Sonohomish |
| 2 | Spain | Jan. 31 | Canary Island | La Gomera |
| 3 | Italy | Jan. 31 | Rome | Rome |
| 5 | Germany | Jan. 27 | Babaria | Munich |
| 6 | UK | Jan. 31 | Newcastle | York City |
| 7 | China | Dec. 31, 2019 | Hubei | Wuhan |
| 8 | Turkey | Mar. 11 | Ankara | Ankara |
| 9 | Iran | Feb. 19 | Qom Province | Qom |
| 11 | Brazil | Feb. 25 | San Paulo | San Paulo |
| 13 | Canada | Jan. 25 | Toronto | |
| 23 | Japan | Jan. 16 | Kanagawa Prefecture | Kanagawa |
| 24 | Korea | Jan. 20 | Seoul | Gimpo |
| 32 | Mexico | Feb. 28 | Mexico City/Sinaloa | Mexico City |

III. Results

The outbreak of the COVID-19 started around January or February 2020 over the world as summarized in Table 1. The disease was first reported at Wuhan City, Hubei province, China in 31 December 2019. Some countries like Republic of Korea, Taiwan, Chinese Hong Kong etc. were controlled well the disease up to now. They have been executed the massive inspection for the disease. When patients were found at a location, the doctors and expert of the CDC (Center for Disease Control and Prevention) checked the all contacting people of the patients. All infected individuals are quarantined in hospital or some remote places. Some suspected persons are self-quarantined on the reported place and the controllers check frequently by App, internet and phone. However, many countries do not prepare to control and to prevent the disease in the early stage like USA and Japan. The patients in the unprepared countries were incubating the disease in the early stage. Recently the countries suffered the abrupt outbreak and many people had been died.

Table 2. Parameters of SIQR$_K$ model. We obtain the fitting constants *a* and *b*. Then we derived the infection rate $\beta$ and the rate $\eta$ of detection of new cases from the fitting value. We obtain the recovering rate $\gamma$ of quarantined individuals from the daily data of Q and R$_K$. We obtain the basic reproduction number $R_o$. Using the fitting parameters, we solve numerically the SIQR$_K$ model and predict the characteristic times.

| Country | a | b | $\chi^2$ | $\beta$ | $I(0)$ | $R_o$ | $\tau$ |
|---|---|---|---|---|---|---|---|
| USA | 1.13 | 0.33 | 0.99 | 0.462 | 17 | 3.5 | 2.1 |
| Spain | 43.7 | 0.19 | 0.99 | 0.324 | 652 | 2.4 | 3.6 |
| Italy | 93.26 | 0.16 | 0.99 | 0.294 | 1392 | 2.2 | 4.3 |
| Germany | 9.62 | 0.25 | 0.99 | 0.384 | 144 | 2.9 | 2.8 |
| UK | 6.6 | 0.21 | 0.99 | 0.344 | 99 | 2.6 | 3.3 |
| China | 563.38 | 00.23 | 0.93 | 0.364 | 8409 | 2.7 | 4.3 |
| Iran | 138.46 | 0.10 | 0.98 | 0.233 | 2077 | 1.8 | 6.9 |
| Brazil | 22.16 | 0.15 | 0.98 | 0.284 | 331 | 2.1 | 4.6 |
| Canada | 0.65 | 0.23 | 0.99 | 0.364 | 10 | 2.7 | 3.0 |
| Japan | 1.06 | 0.12 | 0.96 | 0.254 | 16 | 1.9 | 5.8 |
| Korea | 90.95 | 0.294 | 0.98 | 0.294 | 1356 | 2.2 | 4.3 |
| Mexico | 12.84 | 0.12 | 0.99 | 0.254 | 192 | 1.9 | 5.8 |

In the reported data of each countries, the active cases are transferring immediately to quarantined cases. Therefore, the active cases correspond to the quarantined case Q. Almost recovering cases is coming from the isolated cases. From the reported data for Q and R$_k$ we can fit (Q+R$_k$) as a function of time at the early stage of disease spreading. The (Q+ R$_k$) is fitted by the exponential function as $g(t) = \frac{a}{b}[e^{bt} - 1]$ at the early stage of the disease spreading (see Appendix). From the obtained fitting parameter $a$ and $b$, we estimated the model parameters such as $a = \eta I(0)$ and $b = \beta - (\alpha + \eta)$ where $I(0)$ is the infected individuals at the initial time. We have to determine four parameters $\alpha, \beta, \eta$, and $I(0)$. We determine the rate $\alpha, \eta$, and $\gamma$ according to the method of reference 11. Let $\varepsilon$ denote the fraction of infectious individual entering Q. There are controvercies for the ratio of the asymptomatic cases in the COVID-19 [15-17]. The reported ratio for several cases is in the range between 5% and 80% of the people testing positive for COVID-19 being asymptomatic. We set the fraction as $\varepsilon = 1/3$ [11-17]. The average incubation time is about 5 days [12,13] and the duration of the milder cases of disease is about 5~6 days [14]. The average time of duration from infection to recovery or death of non-isolated cases are about 10 days, corresponding to a rate of 0.1/day [11]. Therefore, we obtain the rates as $\alpha = (1-\varepsilon) \times 0.1$/day and $\eta = \varepsilon \times 0.2$/day. Finally, we obtain the parameters as $\alpha = \eta = 0.067$/day. Using these parameters and the fitting parameter $a$ and $b$, we obtain the parameter $\beta$ and $I(0)$ from the fitting parameters and the predetermined rates. We summarized the results obtained from the data of each countries.

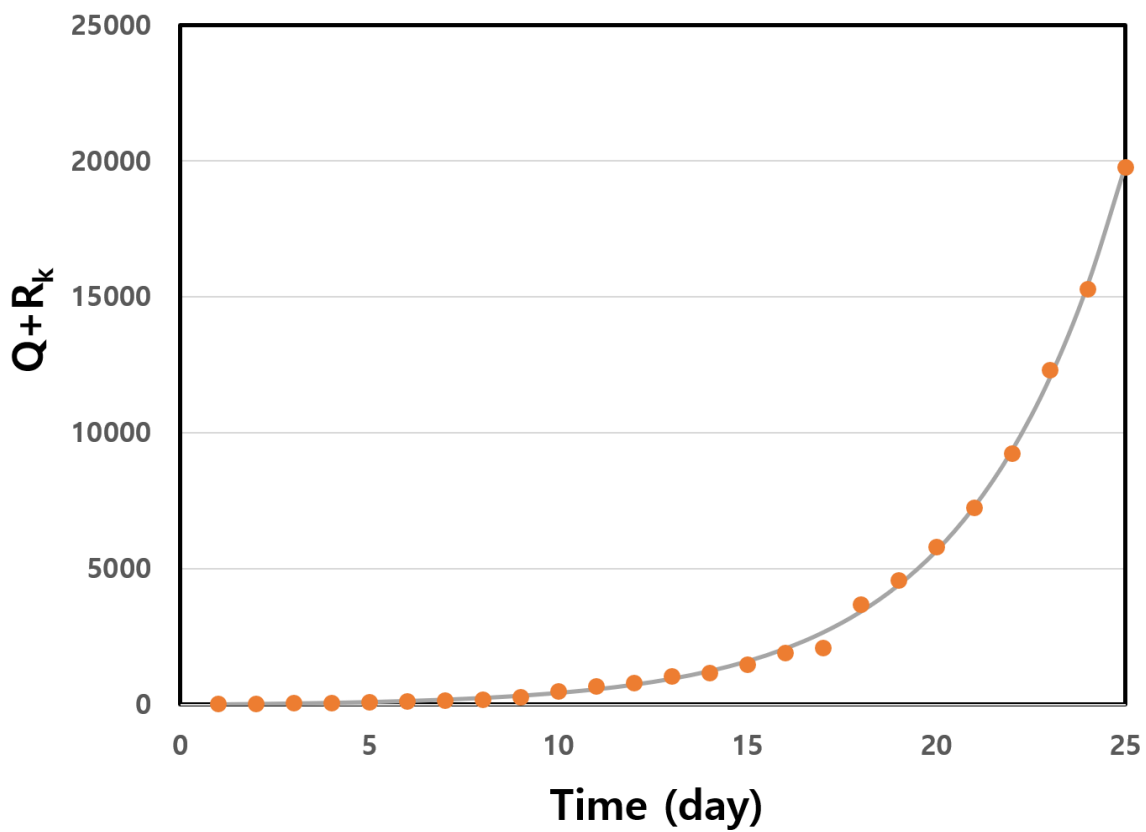

Fig. 1. Q+$R_K$ data were fitted by the nonlinear least square fit at the early time of the outbreak as a function of time for Germany. The solid line is fitting data and solid circles

are the real data. We obtain the fitting parameters as *a*=9.62 and *b*=0.25.

We estimated the infection rate $\beta$ and the initial number of the infected individuals $I(0)$. The symptoms of the COVID-19 are not appearing in many cases. In Fig. 1 we represent the nonlinear square fits of Q+R$_K$ as a function of time at the early stage of the disease spreading in Germany. The early data are well fitted by the exponential function. We give the fitting data for some selected countries in Table 2. We observed that there were large number of initial infected people. The infection rate shows very high value in the range $0.233 \leq \beta \leq 0.462$ for the selected countries. We calculate the basic reproduction numbers of the estimated parameter for the countries. The $R$ of many countries is greater than 2. In particular, the basic reproduction number $R_o$ for the USA shows high value of $R_o = 3.45$. This high value is inducing the large number of infecting people over the states in USA.

We observed the high number of the initial infecting individuals $I(0)$ from the data fitting. In Table 1 we summarized the first official confirmed day of the COVID-19 patient. Because of the incubating period and asymptomatic cases for young health people, we expect that there are many infecting people when the health organization of the country are reporting the first case. In China case we estimated $I(0) = 8409$. The first confirmed time for the virus in China was required for a long time because this disease is a new type of the Coronavirus. For the USA case the initial infecting people is small value as $I(0) = 17$. In US the first patient was found on the state of Washington. However, the late inspection and the delayed quarantine policy by the CDC and the federal government are inducing the huge outbreak in the USA. The South Korea is one of the excellent controlled countries for this disease. In early stage of the outbreak, the initial cases are estimated as $I(0) = 1356$. In South Korea a super-spreader was found at a metropolitan city, Daegu, in Feb. 17, 2020 who attended worships of a church gathering a lot of people. Although the initial infecting people are very big, the Health Organization and Korean CDC are performing the wide range of the inspection, strong policy of the quarantine, and providing the information of the contacting people of the confirmed patient. These strong protecting policies can be preventing the wide spreading of the disease up to now in South Korea.

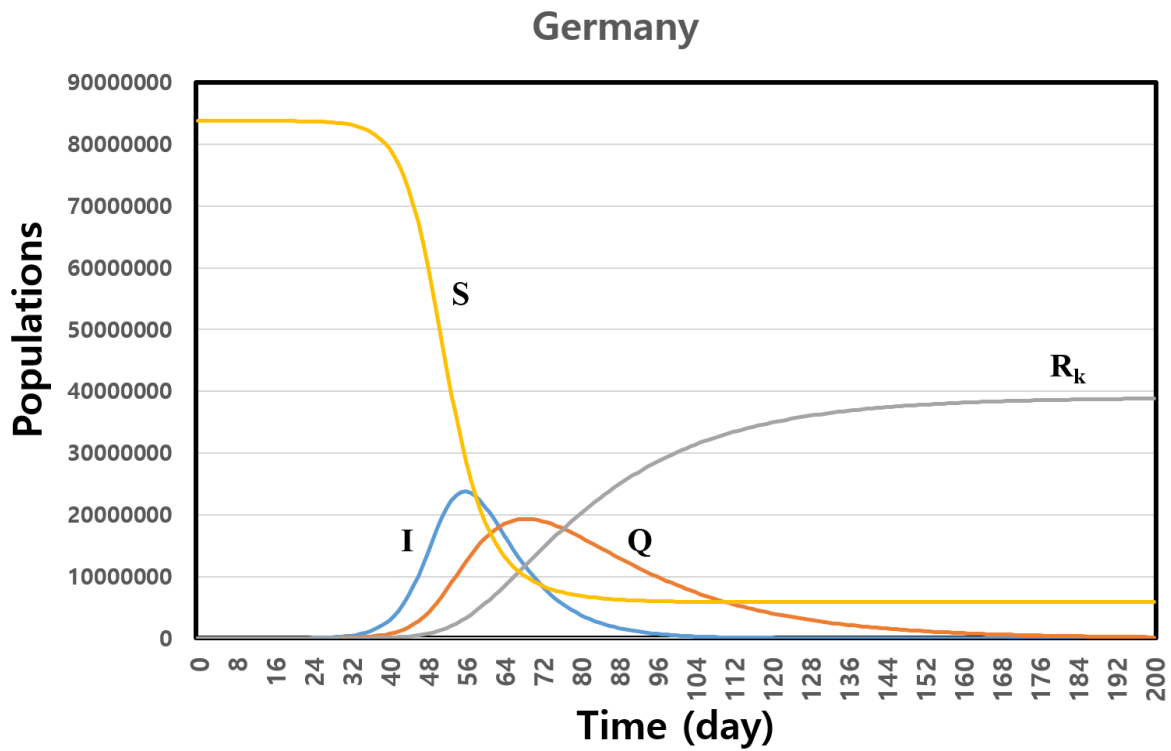

Fig. 2. Prediction of the susceptible (S), infecting (I), quarantined (Q), and confirmed recovered ($R_k$) individuals by the numerical integration using the fitting parameters for Germany. The disease is lasting for 200days after the first occurrence of the patient.

We calculate numerically the SIQR$_K$ model by using the fitting parameters for the countries. Fig. 2 shows the predicted cases of the susceptible (S), infecting (I), quarantined (Q), and officially confirmed recovered ($R_k$) individuals for Germany. The maximum of the infecting people is around 50days when the government are enforcing the quarantine of the infecting person. Of course, the maximum time and the lasting time of the disease depend on the fitting parameters and the initial infecting people. For Germany, the disease is in the calm state after 200days. We require about 6 months to terminate the disease according to our model. We observed that the asymptomatic recovered populations $R_a = N - (S + I + Q + R_k)$ are dramatically increasing after the maximum time of the infection as shown in Fig. 2. When we predict the evolution of the disease by some model, we required to use the confirmed data set such as the active cases, recovered cases, and death cases.

IV. Conclusions

We consider an epidemic spreading model, SIQR$_K$ model. In this model we include the dynamic equation for the quarantined individuals. We estimated the parameters of the dynamic evolution equation from the sum of the quarantined cases and the recovered cases. We obtained the parameters by the nonlinear least square fits by using the reported data set. It is very important that we consider the asymptomatic individuals when we predict the dynamic evolution of the disease by some model. The observed high value of the basic reproduction number indicated the huge pandemic of the disease all over the world. We predict that the maximum time of the infection is around 50 days or two months. The disease should be lasting about six months when we have quarantined the infecting individuals.

Appendix

Let's solve the SIQR$_K$ model. In the early phase of the disease spreading we expect that the susceptible population is similar to the total population $S/N \approx 1$. Therefore, we can write the infection dynamic equation such as [10,11]

$$\frac{dI(t)}{dt} = [\beta - (\alpha + \eta)]I(t). \tag{A1}$$

By integrating this equation with the initial condition $I(0)$, we obtain the solution as

$$I(t) = I(0)e^{[\beta-(\alpha+\eta)]t}. \tag{A2}$$

The reproduction number $R_o$ is given by

$$R_o = \frac{\beta}{\alpha+\eta}. \tag{A3}$$

In the COVID-19, the reproduction number is bigger than one. The disease can spread easily by the contact process between individuals. The double time $\tau$ is given by $\tau = \frac{ln2}{[\beta-(\alpha+\eta)]} = \frac{ln2}{(\alpha+\eta)(R_o-1)}$. The infection rate $\beta$ and the rate $\eta$ of detection of new cases can be derived from the time evolution of the early infection. Adding equation (3) and (4) we obtain a quantity such as

$$\frac{d(Q+R_k)}{dt} = \eta I(t). \tag{A4}$$

Therefore, we obtain the sum of quarantined cases and recovered cases as

$$(Q + R_k)(t) = \frac{\eta I(0)}{\beta-(\alpha+\eta)}\{e^{[\beta-(\alpha+\eta)]t} - 1\}. \tag{A5}$$

We calculate the recovering rate $\gamma$ obtained by the data set. The recovering rate is given by $\gamma = (R_{k_i} - R_{k_{i-1}})/Q_{i-1}$. The value of the recovering rate depends on the time at the early stage and converge to a constant value. We obtained the recovering rate as $\gamma = 0.036$/day.

Acknowledgements

This work was supported by a National Research Foundation of Korea (NRF) grant funded by the Korean Government (Grant No. NRF-2020R1A2C1005334).